\documentclass{article}
\usepackage{spconf,amsmath,graphicx,hyperref}
\usepackage{extra_styles}
\usepackage{booktabs}
\usepackage{graphicx}
\usepackage[caption=false,font=normalsize,labelfont=sf,textfont=sf]{subfig}
\usepackage{threeparttable}
\usepackage{cite}
\usepackage{caption}

\title{LP-CFM: Perceptual Invariance-Aware Conditional Flow Matching\\for Speech Modeling}
\name{Doyeop Kwak, Youngjoon Jang, Joon Son Chung}
\address{Korea Advanced Institute of Science and Technology, South Korea}
\begin{document}
%
\maketitle
\begin{abstract}
The goal of this paper is to provide a new perspective on speech modeling by incorporating perceptual invariances such as amplitude scaling and temporal shifts. Conventional generative formulations often treat each dataset sample as a fixed representative of the target distribution. From a generative standpoint, however, such samples are only one among many perceptually equivalent variants within the true speech distribution. To address this, we propose Linear Projection Conditional Flow Matching (LP-CFM), which models targets as projection-aligned elongated Gaussians along perceptually equivalent variants. We further introduce Vector Calibrated Sampling (VCS) to keep the sampling process aligned with the line-projection path. In neural vocoding experiments across model sizes, data scales, and sampling steps, the proposed approach consistently improves over the conventional optimal transport CFM, with particularly strong gains in low-resource and few-step scenarios. These results highlight the potential of LP-CFM and VCS to provide more robust and perceptually grounded generative modeling of speech.
\end{abstract}

\section{Introduction}
Recent advances in speech generation have identified conditional flow matching (CFM)~\cite{lipmanflow} as a powerful alternative to diffusion-based models. CFM learns a time-dependent vector field that gradually transports samples from a simple source distribution to the complex target data distribution, achieving strong performance in various speech modeling tasks such as speech synthesis, enhancement, and separation~\cite{leeperiodwave,luo2025wavefm,mehta2024matcha,jang2024faces,lee2025flowse,liugenerative}.

From the generative perspective, human auditory perception is generally robust to global amplitude scaling and small temporal shifts. In practice, two waveforms that differ only in loudness or slight temporal alignment are often perceived as perceptually identical~\cite{le2019sdr,zhang2024unrestricted,kwak2025ednet}. This property has already been exploited in several speech-related tasks. For instance, the scale-invariant signal-to-distortion ratio (SI-SDR)~\cite{le2019sdr} is widely adopted as an objective in speech separation for its robustness to amplitude variations~\cite{luo2018tasnet,luo2019conv,subakan2021attention, zhao2023mossformer,zhao2024mossformer2}. Similarly, phase shift-invariant training (PSIT)~\cite{kwak2025ednet} has been shown to enhance both the performance and training stability of speech enhancement models by relaxing strict temporal alignment. In contrast, conventional generative formulations, including CFM, are not inherently designed with such flexibility. They typically enforce learning a single instance from the dataset and penalize any deviation, even when the alternative outputs are perceptually equivalent. This rigid objective could lead to inefficient use of data and model capacity.

\begin{figure}[!t]
    \centering
    \subfloat[\rmfamily\footnotesize OT-CFM]{
        \includegraphics[width=0.22\textwidth]{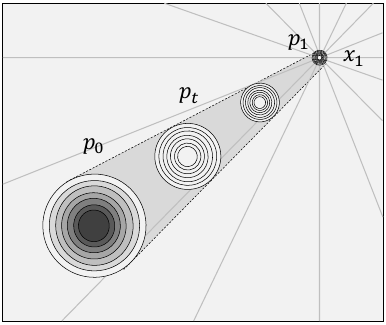}
    } 
    \hfill
    \subfloat[\rmfamily\footnotesize LP-CFM]{
        \includegraphics[width=0.22\textwidth]{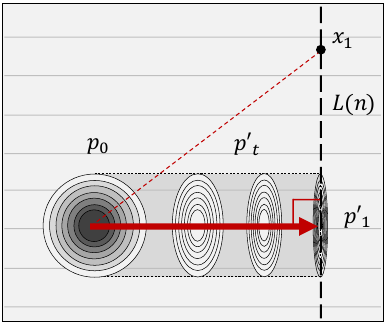}
    }
    \hfill
    \vspace{-2mm}
    \captionsetup{font=small}
    \caption{Conceptual illustration of OT-CFM and LP-CFM: (a) OT-CFM models a spherical (isotropic) Gaussian distribution around target sample $x_1$, whereas (b) LP-CFM places mass along the shortest projection path toward a line formed by equivalent variants of the target sample, resulting in an elongated Gaussian distribution.}
    \vspace{-4mm}
    \label{fig:cfm_overview}
\end{figure}

Motivated by these observations, we propose Linear Projection Conditional Flow Matching (LP-CFM), a new formulation of CFM that explicitly incorporates these perceptual invariances. Rather than matching to an isotropic Gaussian centered on a single data point, LP-CFM defines the target as an elongated Gaussian distribution along a line that represents a set of perceptually equivalent targets (e.g., variants differing only in global amplitude or temporal alignment), as illustrated in ~\Fref{fig:cfm_overview}. This design encourages the model to learn a flow that directs samples toward the closest valid point within the equivalence set, instead of forcing convergence to one arbitrary instance. Furthermore, we introduce Vector Calibrated Sampling (VCS), a simple yet effective correction strategy that ensures sampling remains consistent with the projection-based geometry. Together, LP-CFM and VCS enable the model to capture speech distributions in more efficient and perceptually meaningful way.

To validate the effectiveness of our proposed LP-CFM, we conduct various experiments within a controlled neural vocoding setting. Through a comparative analysis against the optimal transport CFM (OT-CFM), we confirm that our approach achieves consistently better outcomes under diverse conditions. This performance gain is especially pronounced in challenging scenarios, such as limited model capacity or a low number of sampling steps. These findings position LP-CFM as a robust and generalizable alternative for speech generation, and more broadly, as a step toward generative models that align more closely with human perceptual structures.

\section{Background}
Flow matching~\cite{lipmanflow} formulates generative modeling as learning a continuous-time flow that maps a simple prior distribution $p_0$ into the data distribution $p_1$.
While the ideal marginal vector field governing the transport of the entire distribution is intractable, conditional flow matching (CFM) provides a tractable, simulation-free objective. CFM works by defining simpler conditional probability paths $p_t(x|x_1)$ for each data point $x_1$ and training vector $v_\theta$ to match the corresponding conditional vector fields $u_t(x|x_1)$.
A powerful and widely-used implementation is optimal transport CFM (OT-CFM)~
\cite{lipmanflow}, which constructs a straight-line probability path between a prior sample $x_0 \sim \mathcal{N}(0,I)$ and a data sample $x_1$. The intermediate path $x_t$ is defined as an interpolation that transforms the prior into a narrow Gaussian $\mathcal{N}(x_1, \sigma_{\min}^2 I)$ centered at the data point:
$$x_t=(1-(1-\sigma_{\min})t)x_0 + tx_1.$$
For a point on this path, the conditional vector field $u_t(x_t|x_1)$ is a time-invariant vector equivalent to the path velocity $\dot{x}_t$:
$$u_t(x_t|x_1) = \dot{x}_t = x_1 - (1-\sigma_{\min})x_0.$$
The training objective is therefore formulated as a regression problem with the following loss function:
\begin{equation}
\mathcal{L_{\text{CFM}}}(\theta) = \mathbb{E}_{t,x_0,x_1}\left[ \big\| v_\theta(x_t, t) - u_t(x|x_1) \big\|^2 \right].
\label{eq:cfm_loss}
\end{equation}

\section{Methodology}
\subsection{Linear Projection CFM (LP-CFM)}
In practical generation tasks, a given target $x_1$ may have multiple equivalent variants that are perceived as indistinguishable in quality. We assume such variants lie on a line
\[
L(n; x_1) = a(x_1) n + b(x_1), \quad n \in \mathbb{R},
\]
where $a(x_1) \in \mathbb{R}^d$ is a direction vector and $b(x_1) \in \mathbb{R}^d$ is an offset. Any point on $L(n; x_1)$ is then considered a valid variant of $x_1$.

\subsubsection{Target distribution construction}

Under this assumption, instead of modeling the target as a isotropic Gaussian centered at $x_1$, we define an elongated Gaussian distribution that concentrates around the line $L(n; x_1)$. Let $p_0 = \mathcal{N}(\mu_0, \Sigma_0)$ be the source distribution. We construct target distribution as
\[
p'_1(x \mid x_1) = \mathcal{N}\big(b + P(\mu_0 - b), \, M\Sigma_0 M^\top\big),
\]
\vspace{-2mm}
\[
\text{where}\quad M = \lambda I + (1-\lambda) P, 
\quad P = \frac{a a^\top}{a^\top a}, 
\quad \lambda \in (0,1].
\]
Here, $P$ is the projection matrix onto the line direction $a$, and $M$ shrinks orthogonal components by factor $\lambda$. Intuitively, this operation translates $p_0$ toward the closest point on $L(n; x_1)$ and compresses variance in the orthogonal subspace, yielding a thin, elongated Gaussian aligned with the line.
We set $p_0$ as $\mathcal{N}(0,I)$ which in turn simplifies to
\[
p'_1(x \mid x_1) = \mathcal{N}(b - Pb, \, MM^\top).
\]

\subsubsection{Conditional path and velocity}
The conditional probability path is defined as the Wasserstein-2 displacement interpolation between $p_0$ and $p'_1$. Since Gaussians are closed under $W_2$ displacement interpolation~\cite{takatsu2011wasserstein}, the intermediate distribution $p'_t$ is Gaussian as well. At the sample level, the interpolation can be written as
\[
x_t = (1-t)x_0 + t\,(b - Pb + Mx_0), 
\quad x_0 \sim p_0,
\]
with target velocity
\[
u_t(x|x_1) = \dot{x}_t  = (b - Pb + Mx_0) - x_0.
\]
The training objective follows the CFM form as \Eref{eq:cfm_loss}.

This formulation naturally includes OT-CFM as a special case: when $\lambda = \sigma_{\min}, \, b = x_1$, and the line is undefined ($a = 0 \Rightarrow P = 0$), 
the formulation becomes identical to the isotropic Gaussian target of OT-CFM.
Our method can therefore be viewed as a more general formulation that adapts the covariance structure to reflect equivalence classes of data along $L(n; x_1)$.


\subsection{Application on Speech Modeling}
Theoretically, LP-CFM can be applied to any kind of generation task if its variants can be expressed as a line equation $L(n) = an + b$. To provide a concrete instantiation of our general theory, this paper focuses on the task of speech modeling. We formulate the proposed equations for this specific domain by leveraging the core properties of the short-time Fourier transform (STFT).

\subsubsection{Scaling property}
When a signal is scaled in amplitude by a factor $s$, its magnitude spectrogram becomes $X_{\text{mag},y} = |s| X_{\text{mag}}$. Taking the logarithm of this equation yields an additive relationship: 
\[\log X_{\text{mag},y} = \log X_{\text{mag}} + \log|s|.\] 
By setting the variant parameter $n = \log|s| \in \mathbb{R}$, this forms the line equation $L(n) = \log X_{\text{mag}} + n$, which corresponds to the line with a slope of $a=1$ and an offset of $b=\log X_{\text{mag}}$.

\subsubsection{Shifting property}
When a signal is shifted in time of $\tau$, the phase spectrogram $X_{\text{pha}}$ is modifies as follows:
\vspace{-1mm}
\[
X_{\text{pha},y}[k] = X_{\text{pha}}[k] - \frac{2\pi k}{N} \tau,
\]
where $k$ is the frequency-bin index and $N$ is the FFT size. This expression is inherently a line equation. By defining $n=\tau$ and a constant vector $\kappa[k] = 2\pi k / N$, the equation takes the form $L(n) = X_{\text{pha}} - n\kappa$. This corresponds to a line with a slope of $a=-\kappa$ and an offset of $b=X_{\text{pha}}$.
Through these constructions, both log-magnitude and phase spectrograms admit linear variant sets. This derivation can also be applied to broader domains, such as log-mel spectrograms, which share the same scaling property.

\subsection{Vector Calibrated Sampling (VCS)}
In LP-CFM, the target velocity $u_t$ is by definition orthogonal to its corresponding target line. However, the predicted vector $v$ may contain small, erroneous components parallel to this line due to prediction errors.
To address this, we propose Vector Calibrated Sampling (VCS), a simple correction applied during inference to enforce this geometric constraint. VCS removes the erroneous component of the predicted vector $v$ that is parallel to the target line, while preserving the vector's original magnitude:
\vspace{-1mm}
\[
v' = \frac{\|v\|}{\|(I - P)v\|}(I - P)v.
\]
This operation is feasible in our speech application because the line slopes are known constants ($a=1$ for log-magnitude and $a=-\kappa$ for phase). The purpose of VCS is not to significantly boost performance, but to act as a safeguard that ensures the sampling process remains consistent with the geometric properties of the LP-CFM framework.
\section{Experiments}
\subsection{Experimental Setup}
We evaluate our method on a neural vocoding setup, converting mel-spectrograms into waveforms. This task serves as a controlled testbed for modeling both magnitude and phase, allowing for a relative comparison against OT-CFM across varying conditions to isolate the contributions of LP-CFM.

\newpara{Model architecture.} 
To control for architectural factors, we simplify the model design. The mel encoder consists of 1D convolution with kernel size of 7 followed by a single ConvNeXt V2~\cite{woo2023convnext} block, which maps mel bins to the STFT frequency dimension. The encoded mel features are channel-wise concatenated with the input of a flow matching decoder to predict vectors for both magnitude and phase spectrograms. The decoder is a minimally modified open-source 2D UNet backbone\footnote{\href{https://huggingface.co/docs/diffusers/api/models/unet2d}{https://huggingface.co/docs/diffusers/api/models/unet2d}}, featuring one ResNet block per scale, three scales in total with no attention modules. We build three model sizes with channel configurations of [16,32,64], [32,64,128], and [64,128,256], using group normalization with 2, 4, and 8 groups, respectively. The decoded magnitude and phase spectrograms are converted to a waveform via an inverse STFT.

\newpara{Training details.} 
All experiments use the single-speaker LJ Speech~\cite{ljspeech17} dataset. Following prior work~\cite{kong2020hifi}, mel-spectrograms and target STFTs are extracted using a 1024-point FFT, a 256-sample hop size, and 80 mel bins (0--8 kHz). We use a train-validation split of 12,950 and 150 samples~\cite{kong2020hifi}.
To ensure a fair comparison, both LP-CFM and OT-CFM are trained with identical settings and a fixed random seed. We set $\lambda$ as $1\times10^{-4}$, matching the $\sigma_{min}$ value used in OT-CFM. Models are trained for 500 epochs on a single RTX 4090 GPU with a batch size of 16. We use the AdamW optimizer with a betas of (0.9, 0.99), learning rate of $5\times 10^{-4}$, decayed exponentially by 0.99 per epoch.

\newpara{Evaluation metrics.} 
We report performance using standard vocoder metrics: multi-resolution STFT (M-STFT)~\cite{yamamoto2020parallel}, PESQ~\cite{rix2001perceptual}, mel-cepstral distance (MCD)~\cite{kubichek1993mel}, periodicity error and V/UV F1~\cite{morrisonchunked}, along with UTMOS~\cite{saeki2022utmos} as an automated proxy for subjective quality. For sampling, we utilize first-order Euler ODE-solver with sampling step of 6 as a default. Since LP-CFM can produce outputs with various scales, all target and predicted waveforms for both methods are peak-normalized to 0.95 before evaluation.

\begin{table}[t!]
    \centering
    \captionsetup{font=small}
    \caption{Results under differenc model sizes.}
    \vspace{-2mm}
    \resizebox{0.99\linewidth}{!}{
    \begin{tabular}{c c|ccccc|c}
        \toprule
        \textbf{Model} & \textbf{Method} & \textbf{M-STFT$\downarrow$} & \textbf{PESQ$\uparrow$} & \textbf{MCD$\downarrow$} & \textbf{Period$\downarrow$} & \textbf{V/UV F1$\uparrow$}  & \textbf{UTMOS$\uparrow$} \\
        \midrule
        \multirow{2}{*}{UNet-16} 
            & OT & 1.0399 & 3.743 & 2.223 & 0.1108 & 0.9596  & 2.8715 \\
            & LP & \textbf{1.0253} & \textbf{3.858} & \textbf{2.174} & \textbf{0.1050} & \textbf{0.9614}  & \textbf{3.0153} \\
        \midrule
        \multirow{2}{*}{UNet-32} 
            & OT & 0.9917 & 4.011 & 2.048 & 0.0908 & 0.9655 & 3.2254 \\
            & LP & \textbf{0.9848} & \textbf{4.097} & \textbf{2.018} & \textbf{0.0881} & \textbf{0.9665}  & \textbf{3.2647} \\
        \midrule
        \multirow{2}{*}{UNet-64} 
            & OT & 0.9670 & 4.180 & 1.975 & 0.0801 & 0.9704  & 3.3900 \\
            & LP & \textbf{0.9631} & \textbf{4.191} & \textbf{1.942} & \textbf{0.0772} & \textbf{0.9709}  & \textbf{3.4231} \\
        \bottomrule
    \end{tabular}
    }
    \label{tab:model_size}
    \vspace{-4mm}
\end{table}

\subsection{Results and Analysis}
\newpara{Impact of model size.}
The analysis begins with model capacity, examining how it influences the relative behavior between LP-CFM and OT-CFM. As shown in \Tref{tab:model_size}, LP-CFM provides consistent gains across multiple architectures. The improvements are particularly notable when the model size is small (e.g., UNet-16), and the gap narrows for larger models. This performance trend can be attributed to line-projection geometry of LP-CFM: by targeting the closest point on a line of valid variants rather than path converging to a single fixed point, the transport path length and variability are reduced. This property may ease the optimization, especially for models with limited capacity.

\begin{table}[t!]
    \centering
    \captionsetup{font=small}
    \caption{Results under different dataset sizes on UNet-32.}
    \vspace{-2mm}
    \resizebox{0.99\linewidth}{!}{
    \begin{tabular}{c c|ccccc|c}
        \toprule
        \textbf{Trainset} & \textbf{Method} & \textbf{M-STFT$\downarrow$} & \textbf{PESQ$\uparrow$} & \textbf{MCD$\downarrow$} & \textbf{Period$\downarrow$} & \textbf{V/UV F1$\uparrow$}  & \textbf{UTMOS$\uparrow$} \\
        \midrule
        \multirow{2}{*}{LJ - 33\%} 
            & OT & 1.0176 & 3.929 & 2.124 & 0.0992 & 0.9618  & 3.1118 \\
            & LP & \textbf{1.0153} & \textbf{3.975} & \textbf{2.101} & \textbf{0.0976} & \textbf{0.9634}  & \textbf{3.1501} \\
        \midrule
        \multirow{2}{*}{LJ - 66\%} 
            & OT & 1.0047 & 3.994 & 2.051& 0.0941 & 0.9646 & 3.1718 \\
            & LP & \textbf{0.9968} & \textbf{4.071} & \textbf{2.037} & \textbf{0.0902} & \textbf{0.9669}  & \textbf{3.2416} \\
        \midrule
        \multirow{2}{*}{LJ - 100\%} 
            & OT & 0.9917 & 4.011 & 2.048 & 0.0908 & 0.9655  & 3.2254 \\
            & LP & \textbf{0.9848} & \textbf{4.097} & \textbf{2.018} & \textbf{0.0881} & \textbf{0.9665}  & \textbf{3.2647} \\
        \bottomrule
    \end{tabular}
    }
    \label{tab:subset}
    \vspace{-2mm}
\end{table}
\newpara{Data efficiency.}
To compare data efficiency between the two methods, we train models on randomly sampled subsets of LJSpeech and compare their performance. 
\Tref{tab:subset} demonstrates that LP-CFM consistently surpasses OT-CFM even under limited data scenarios. 
For example, training with only 66\% of the data still yields higher performance than OT-CFM trained on the full dataset across most metrics. Since LP-CFM constructs its target distribution by capturing 
a set of multiple variants with a single elongated Gaussian, the model can leverage a richer and more diverse set of data instance than the dataset alone provides. This approach resembles data augmentation, but with a crucial distinction: instead of exposing on arbitrary variants, LP-CFM dynamically steers the learning process toward the closest variant under the current flow.

\begin{figure}[t!]
   \centering
   \includegraphics[width=0.95\linewidth]{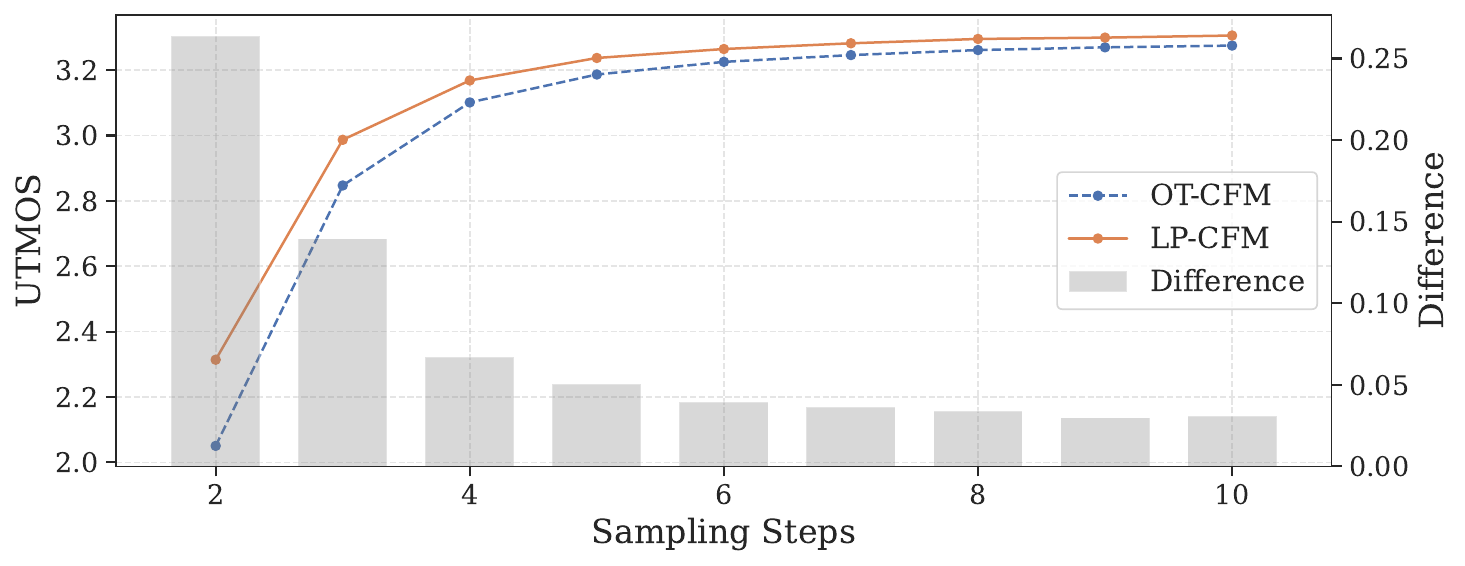}
   \vspace{-3mm}
   \captionsetup{font=small}
   \caption{Step-wise UTMOS comparison between OT-CFM and LP-CFM on UNet-32. UTMOS scores are shown as line (left axis), and the bar represents their score difference (right axis).}
   \label{fig:step_utmos}
\vspace{-5mm}
\end{figure}
\newpara{Sampling efficiency.}
Building on the above findings, we next examine how the line-projection geometry affects inference behavior. We evaluate UTMOS performance under varying step budgets. As shown in \Fref{fig:step_utmos}, LP-CFM consistently achieves higher scores across different numbers of steps, with its advantage most evident in few-step regimes where approximation errors tend to accumulate. These results suggest that the proposed line-projection geometry—which yields shorter and more consistent transport paths—not only facilitates optimization and improves data efficiency, but also proves effective in sampling, particularly in low-step settings where error accumulation is a concern.

\begin{table}[t]
    \centering
    \captionsetup{font=small}
    \caption{Results of CMOS test between OT-CFM and LP-CFM on representative scenarios with 95\% confidence interval. Higher scores indicate stronger preference for LP-CFM.}
    \vspace{-2mm}
    \resizebox{0.5\linewidth}{!}{
    \begin{tabular}{l| c }
    \toprule
    \textbf{Scenario} & \textbf{CMOS $\uparrow$}  \\
    \midrule
    UNet-32, 6 steps        &  0.12$\pm$0.09\\
    UNet-16, 6 steps        &  \textbf{0.46$\pm$0.10}\\ 
    UNet-32, 33\% data, 6 steps &  0.17$\pm$0.10\\
    UNet-32, 3 steps        &  0.35$\pm$0.12 \\
    \bottomrule
    \end{tabular}
    }
    \vspace{-2mm}
    \label{tab:cmos}
\end{table}
\newpara{Subjective evaluation.}
To examine how the observed objective gains translate into perceptual quality, we conducted a comparative mean opinion score (CMOS) evaluation on 15 randomly chosen validation samples, each rated by 25 listeners across four representative scenarios. We verify the results with one-sample t-tests against zero, confirming that all results are statistically significant (p-value $<$ 0.05). As shown in \Tref{tab:cmos}, listeners express a clear preference for LP-CFM in small-model and few-step sampling conditions. In the other scenarios, LP-CFM also receive consistently positive ratings, with relatively higher preference under the low-data setting. Taken together, these results indicate that the perceptual advantages of LP-CFM are consistent with its objective improvements.

\begin{table}[t!]
    \centering
    \captionsetup{font=small}
    \caption{Results of ablation study on UNet-32.}
    \vspace{-2mm}
    \resizebox{0.99\linewidth}{!}{
    \begin{tabular}{ccc|c|ccccc|c}
        \toprule
        \multirow{2}{*}{} & \multicolumn{2}{c|}{\textbf{Method}} & \multirow{2}{*}{\textbf{VCS}} 
            & \multirow{2}{*}{\textbf{M-STFT$\downarrow$}} 
            & \multirow{2}{*}{\textbf{PESQ$\uparrow$}} 
            & \multirow{2}{*}{\textbf{MCD$\downarrow$}} 
            & \multirow{2}{*}{\textbf{Period$\downarrow$}} 
            & \multirow{2}{*}{\textbf{V/UV F1$\uparrow$}} 
            & \multirow{2}{*}{\textbf{UTMOS$\uparrow$}} \\
        & \textbf{Mag.} & \textbf{Pha.} &  &  &  &  &  &  &  \\
        \midrule
        \textit{(1)} & OT & OT & x & 0.9917 & 4.011 & 2.048 & 0.0908 & 0.9655  & 3.2254 \\
        \textit{(2)} & OT & OT & o & 5.4160 & 1.102 & 11.138 & 0.6437 & 0.0058  & 1.6226 \\
        \midrule
        \textit{(3)} & OT & LP & x & 0.9935 & 4.016 & 2.030 & 0.0909 & 0.9658  & 3.2263 \\
        \textit{(4)} & LP & OT & x & 0.9856 & 4.088 & 2.022 & 0.0880 & \textbf{0.9665}   & 3.2550 \\
        \midrule
        \textit{(5)} & LP & LP & x & 0.9859 & 4.094 & 2.019 & \textbf{0.0879} & \textbf{0.9665}  & 3.2627 \\
        \textit{(6)} & LP & LP & o & \textbf{0.9848} & \textbf{4.097} & \textbf{2.018} & 0.0881 & \textbf{0.9665}   & \textbf{3.2647} \\
        \bottomrule
    \end{tabular}
    }
    \vspace{-4mm}
    \label{tab:ablation}
\end{table}

\newpara{Ablation study.}
To disentangle the contributions of each component, we evaluate LP-CFM when applied separately to magnitude (row 4) and phase (row 3), in comparison with OT-CFM applied to both (row 1). As shown in \Tref{tab:ablation}, applying it to the magnitude yielded the dominant improvements, while phase-only application resulted in smaller gains. This outcome can be attributed to the dominant role of the magnitude on speech quality, as well as the inherent complexity of phase modeling. Ultimately, applying LP-CFM to both components produced the most balanced performance (row 5).

We also examine the effect of VCS. When combined with LP-CFM, VCS behaved as an intended safeguard: it neither boosted performance nor harmed it, yielding comparable or slightly higher scores (row 6). In contrast, applying VCS to OT-CFM significantly degraded performance (row 2), which is expected since it does not assume projection-aligned trajectories. This contrast provides indirect evidence that LP-CFM has indeed learned the intended projection-aligned paths. Since VCS explicitly removes the parallel component, a model not following such trajectories would be expected to suffer the same degradation observed in OT-CFM.

\section{Conclusion}
In this work, we introduced LP-CFM, a perceptual invariance–aware refinement of conditional flow matching that aligns training with perceptual equivalence in speech. As a proof of concept, we evaluated the proposed method in a controlled neural vocoding setup, where it delivered consistent gains over OT-CFM across diverse conditions. Its advantages were most pronounced in resource-constrained scenarios, including limited model capacity, data scarcity, and few-step sampling—conditions often encountered in practical applications. 
We expect that LP-CFM will serve as a foundation for more perceptually informed generative speech models and inspire further exploration of invariance-aware modelings.

%
%
\bibliographystyle{IEEEbib}
\bibliography{shortstrings,mybib}

\end{document}